\input phyzzx
\newcount\mongocount
\mongocount=1
\def\Figure#1#2#3{
% \boxit{
      \vbox to #3in{\hsize=#2in
        \vfil
%\special{ps::[begin]
%          save 10 dict begin /Figure exch def
%          currentpoint translate
%          /showpage {} def
%        }
%        \special{ps: plotfile #1}
         \includegraphics{#1}
%        \special{ps::[end]
%        clear Figure end restore
%        }
    }
% }
}
\def\figcap#1#2{
\vtop{\tenpoint\singlespace
\hsize=#1in\smallskip\noindent Figure\ \ \the\mongocount.\ \  #2
\global\advance\mongocount by 1\bigskip}}
\def\mongofigure#1#2#3#4#5{\centerline{\Figure{#1}{#2}{#3}
\figcap{#4}{#5}}}

\hoffset=0.375in
\overfullrule=0pt

\def\kms{\,{\rm km}\,{\rm s}^{-1}}
\def\dol{d_{\rm {ol}}}
\def\dos{d_{\rm {os}}}
\def\dls{d_{\rm {ls}}}
\def\re{r_{\rm e}}
\def\te{t_{\rm e}}
\def\dsat{d_{\rm{sat}}}
\def\vsat{v_{\rm{sat}}}
\def\ve{v_{\oplus}}
\twelvepoint
\font\bigfont=cmr17
\bigskip
\centerline{}
\centerline{}
%\centerline{}
\centerline{\bigfont Satellite Parallaxes of Lensing} 
\centerline{\bigfont Events Towards the Galactic Bulge} 
 
\bigskip
\centerline{\bf B. Scott Gaudi}
\smallskip
\centerline{and}
\smallskip
\centerline{\bf Andrew Gould$^{*}$}\foot{Alfred P. Sloan Foundation Fellow}
\smallskip
\centerline{Dept.\ of Astronomy, The Ohio State University, Columbus, OH 43210}
\smallskip
\centerline{e-mail gaudi@payne.mps.ohio-state.edu}
\centerline{e-mail gould@payne.mps.ohio-state.edu}
\bigskip
\smallskip
%\doublespace
\singlespace
\centerline{\bf Abstract}

In order to understand the nature of the lenses
that generate microlensing events, one would like to measure their mass,
distance, and velocity.  Unfortunately, current 
microlensing experiments measure only one parameter of the events,
the characteristic timescale, which is a combination of the 
underlying physical 
parameters.  Other methods are required
to extract additional information.  Parallax measurements using
a satellite in an Earth-like orbit yield the projected
velocity of the lens: ${\bf{\tilde v}} ={\bf{v}}/(1-z)$, where 
${\bf{v}}$ is the transverse velocity (speed and direction)
of the lens relative to the
Earth-source line of sight, and $z$ is the ratio
of the distances to the lens and the source.  A measurement of
${\bf{\tilde v}}$ could distinguish between lenses belonging to the bulge 
and disk populations.   We show that for photometric precisions of 1\% to
2\%, it is possible to measure the projected speed, $\tilde v$,
to an accuracy of 
$\leq 10\%$ for over 70\% of disk lenses and over 60\% of bulge lenses.
For measuring the projected velocity ${\bf{\tilde v}}$,
the percentages are 40\% and {30\%}, respectively.
We find lines of sight $> 2^{\circ}$ away from the ecliptic are preferable,
and an Earth-satellite separation in the range $0.7 {\rm{AU}} - 1.9{\rm{AU}}$
is optimal.  The requirements of the satellite for measuring the projected 
velocities of events towards the bulge are similar to those for measurements
toward the LMC. 

\bigskip
Subject Headings:  gravitational lensing -- Galaxy: stellar content
%\doublespace
%\normalspace

\smallskip
\centerline{submitted to {\it The Astrophysical Journal}: January 9, 1995}
\centerline{Preprint: OSU-TA-1/96}
\endpage
\normalspace
\chapter{Introduction}
%\normalspace
%\doublespace
Four ongoing microlensing searches have detected more than 100 candidate
events, the great majority toward the Galactic bulge. 
The MACHO collaboration (Alcock et al.\ 1995b, 1996b) 
have reported 8 candidate
microlensing events of stars in the Large Magellanic Cloud (LMC), more than
expected by the known luminous stars
of the Milky Way (Bahcall et al.\ 1994; Gould, Bahcall \& 
Flynn 1996) and LMC (Gould 1995b), 
but less than expected from standard spherical halo
models (Paczy{\'n}ski 1986; Griest 1991).  The EROS collaboration (Aubourg
et al.\ 1995, Ansari et al.\ 1996) have 
reported 2 candidate events towards the LMC.
The MACHO (Alcock et al. 1995a, 1996a), 
OGLE (Udalski et al. 1994), and DUO (Alard 1996) collaborations 
have reported more than 100
candidate events of stars towards the galactic bulge.  This is also more
than expected by known stars.  These discrepancies suggest that a revision
of the current models of the structure of the disk, bulge, and halo may be 
in order.

The light curve generated by a microlensing event is described by the
magnification $A(x)=(x^2 +2)/x(x^2 +4)^{1/2}$, where $x(t)$ is the position
of the lens
in the Einstein ring as a function of time, and is given by,
$x(t)=[\omega^2(t-t_0)^2 + \beta^2]^{1/2}$.  Here $t_0$ is the time of maximum 
magnification, $\beta$ is the impact parameter in units of the Einstein
ring radius, and $\omega=\te^{-1}$ is the inverse Einstein ring crossing time.
Thus the magnification curve is fit by three parameters, $\omega, \beta, t_0$.
Of these three parameters, only $\omega$ yields any information about the 
lens itself; $\beta$ and $t_0$ reflect only the geometry of the event.
The timescale is related to the Einstein ring radius and velocity of the 
lens by $\te = \re / v$.  Here $v=|{\bf{v}}|$ is the magnitude of the 
transverse velocity of the lens along the Earth-source line of sight,
$$
{\bf{v}} = {\bf{v}}_{\rm{l}} - \left[{\bf{v}}_
{\rm{s}}{\dol\over{\dos}}+{\bf{v}}_{\rm{o}}
{\dls\over{\dos}}\right],\eqn\transvel
$$
where ${\bf{v}}_{\rm{s}}$, ${\bf{v}}_{\rm{l}}$, and ${\bf{v}}_{\rm{o}}$ 
are the transverse velocities of the source, lens
and observer, respectively, $\dol$ and $\dos$ are the distances
to the lens and source, and $\dls$ is the distance between the lens and the
source.  The Einstein ring radius is defined,
$$
\re^2={4GM\over c^2}\dos z(1-z),\,\,\,\,z={{\dol}\over{\dos}}.\eqn\einrad
$$
Thus the timescale is a combination of the lens's physical characteristics:
$M$, $\dol$, and $v$. 
Because the bulge and disk of the Galaxy are characterized by different 
speed and distance distributions, separate measurements of $v$ and $\dol$
would allow one to determine the component to which the lens belongs.  
Measurement of $M$ would yield the mass spectrum of
lenses in these components.  Unfortunately, a measurement of $\te$ alone
provides no direct information about these characteristics.
Some information can be obtained by considering probability distributions, but
this method is inherently statistical in nature.  Furthermore,
lenses belonging to the Galactic bulge will produce events
with time scales very similar to those produced by lenses belonging to the
Galactic disk.  As a result, measurements of $\te$ alone 
differentiate poorly between these two components.

In order to discriminate between the bulge and disk components,
one must consider methods of extracting more information from each event.  
Several methods have been proposed, including using parallax to measure
the projected velocity of the lens, $\tilde {\bf{v}} \equiv {\bf{v}}/(1-z)$. 
There are two basic methods of acquiring
information from parallax: ground-based measurements and space-based 
measurements.  The projected velocity of one lens has already been measured
using ground-based parallax (Alcock et al. 1995c).  However,  
ground-based parallax is limited to those events 
for which the crossing time is large, $\te \gsim 2\,\rm{months}$;  it is only 
for these events that the motion of the Earth induces measurable
asymmetries in the light curve.  Unfortunately, most of the events that 
have been detected so far have time scales significantly smaller than those
required for ground-based parallax.

Ground-based parallax illustrates a generic problem inherent in most of the
methods that have been proposed for measuring the physical characteristics of
lenses.  These methods are only effective for a small fraction of events.  
Furthermore, those events for which these methods are viable 
typically represent a very biased subsample.  Measurements of proper motions 
are biased towards bulge events, color-shift measurements are biased towards 
bright lenses, and ground-based parallax measurements are biased toward events
with long durations.  In contrast, satellite-based parallax measurements are
effective for a majority of events, 
and are not biased towards a particular class of 
events.

Two basic methods have been suggested of measuring 
the proper motion of the lens, $\mu = v/\dol$.  These are photometric 
(Gould 1994; Nemiroff \& Wickramasinghe 1994; Witt \& Mao 1995; Witt 1995;
Loeb \& Sasselov 1995; Gould \& Welch 1996) and spectroscopic (Maoz \& Gould
1994).  Ideally one would like to combine measurements of both the 
proper motion and the parallax to determine both $\tilde v$ and $\mu$.  These
two parameters, along with the timescale, completely determine $M$, $\dol$, and
$v$ (Gould 1996).  (Combined measurements of the proper motion and parallax can be used
to determine the true velocity of the lens only if
the projected velocity, ${\bf{\tilde v}}$,  is determined, 
and not just the projected speed
$\tilde v$.)\ \ 
As we discuss below, parallax measurements can be used to determine 
$\tilde {\bf{v}}$ in a significant fraction of events.
Thus the mass spectrum, physical distribution, and velocity distribution
of lenses could be determined if both the parallax and proper motion of events
could be measured.  Unfortunately, while
optical interferometers may be able to measure proper motions for
many events in the relatively near future, 
events for which proper motion can currently be measured are
rare.

\FIG\one{
The geometry of the orbits of the Earth and the satellite.  Note in 
particular the angle $\psi$, defined as the angle such that 
when $\psi = 0$, the projected displacement vector ${\bf{r}}$ is 
most closely aligned with the line of sight.
}

\FIG\two{
Degeneracy Breaking Fraction (DBF) as a function of $\psi$ for
disk-bulge events (solid bold curve) and bulge-bulge events (dashed curve).  All other
parameters are fixed at $n_{obs} = 20$, $(\sigma_E,\sigma_S)=(1\%,2\%)$, 
$M=0.3M_{\odot}$, $d_{sat}=60^{\circ}$, $\beta_{max}=0.7$, and
$\alpha=84^{\circ}$.  The DBF is the fraction of events for which
the allowed solutions differ by $< 20 \%$ and each of these has intrinsic
error $< 10 \%$.  The lower solid curve is the DBF as a function of
$\psi$ when allowed solutions differ by $< 5\%$, and each has intrinsic
error $< 10 \%$.
}

\FIG\three{
DBF as a function of $\psi$ and
$\beta$ for disk-bulge events. 
The labeled bold contour represents a DBF of 90\%.  Other bold contours
represent DBFs of 50\% and 10\%.  Intermediate contours are spaced at
intervals of 10\%.  All other parameters are held fixed at our fiducial
parameters, as in Fig.\ {\two}.
}

\FIG\four{
DBF as a function of $\psi$ and
$\log{M/M_{\odot}}$ for disk-bulge events. 
Contours are the same as those in Fig.\ {\three}. 
}

\FIG\five{
DBF as a function of $\psi$ and
${\rm{B}}$ for disk-bulge events. 
Contours are the same as those in Fig.\ {\three}.
}

It is not strictly necessary, however, to measure both $\mu$ and $\tilde v$ in
order to distinguish between component populations of lenses.  Han \& Gould
(1995) demonstrated that, for events towards the galactic bulge, the typical
values of $\tilde v$ are reasonably well separated.  Thus a measurement of
$\tilde v$ alone, gathered from parallax information, can be used to determine
the component to which 
the lens belongs.  Furthermore, parallax measurements can 
often be used to determine not only $\tilde v$, but also
the projected velocity of the lens, $\tilde {\bf{v}}$.  Information about
$\tilde {\bf{v}}$ is even more useful in distinguishing between bulge and disk 
populations (Han \& Gould 1995).

Space-based parallax is the
most promising method of extracting information about lenses.   
Gould (1995a)
demonstrated the basic method of using information obtained by a satellite
in an Earth-like orbit to measure ${\bf{\tilde v}}$, 
and also gave a rough estimate of the photometric precision needed.
Due to parallax, the light curve of a particular microlensing
event will be different as seen from the Earth and
the satellite.  This difference is given by
the vector displacement of the lens in the Einstein ring, 
${\Delta\bf{x}}  \equiv (\omega \Delta t,\Delta \beta)$.  Here $\Delta t$ is
the difference between the time of maximum magnification for the Earth and 
satellite, $\Delta t = t_0'-t_0$, and $\Delta \beta$ is the difference between
the impact parameters, $\Delta \beta = \beta' - \beta$.  The vector 
displacement is related to 
the Earth-satellite separation by 
$$
{\Delta\bf{x}}={{\bf r}\over{\tilde \re}},\eqn\delx
$$ 
where ${\bf r}$ is the projection of the Earth-satellite separation vector
$\bf R$, onto the plane of the sky (perpendicular to the source vector 
${\bf \hat s}$), and $\tilde \re$ is the projected Einstein ring radius,
$$
\tilde \re = {{\re}\over{1-z}}.\eqn\redein
$$
Thus by measuring $t_0$ and $\beta$ for the
Earth and the satellite, one can determine the projected Einstein ring radius,
and this, combined with the timescale $\te$, can be used to calculate the
projected velocity of the lens, ${\bf{\tilde v}}$.

Complications arise from the fact that there is a four-fold degeneracy
inherent in the determination of ${\bf \tilde v}$, including a 
two-fold degeneracy in the projected speed, $\tilde v$.  
These degenerate cases
arise from the fact the one does not know {\it{a priori}} which side
of the lens-source line the Earth and the satellite are on.  
Thus $\Delta \beta$ can have two magnitudes: $\Delta \beta _{\pm} = |\beta'
\pm \beta|$, and two distinct signs: $\pm\Delta \beta _{\pm} = \pm\ | \beta'
\pm \beta|$, for a total of four degenerate cases.  
This degeneracy can, in principle, be broken by measuring the
small difference in inverse time scales, $\Delta \omega = \omega' - \omega$, 
caused by the relative velocity of the Earth
and the satellite (Gould 1995a).

Launching a satellite for observing microlensing
would be a major undertaking.  It is often suggested that the most economical approach would
be to attach the telescope to an already planned mission.  It turns out,
however, that the stability requirements for observations of microlensing events
are such that it would be more economical to launch a dedicated satellite.  
The primary concern, therefore, is that such a dedicated
satellite would be sufficiently 
versatile and beneficial to justify the expense.  We have 
already discussed what could be gained from a measurement of ${\bf{\tilde v}}$ of
lenses, but it remains to be shown that a satellite with reasonable 
specifications will actually enable one to measure ${\bf{\tilde v}}$, and
whether these measurements could be made for a large class of microlensing events.
There is the further consideration that there are two distinct lines of sight of
microlensing observations, the LMC and the galactic bulge.  The times of year when
these two lines of sight can be observed from the ground are complementary, and one would like
one satellite to be able to make measurements for both cases.  Since the two cases are
markedly different in the nature of events that are expected, the brightness of
the source stars being observed, and in the geometry of the parallax measurements,
it is not obvious that a satellite with specifications suited to
LMC parallax measurements will also be able to make parallax measurements towards
the galactic bulge.  

It is interesting to ask what the required lifetime of a parallax satellite would be to
achieve a reasonable statistical sample of microlensing events. 
Gould (1995a) estimated that $\sim 100$ giant star events will be seen towards the bulge
during each $\sim 6$ month bulge season.  To observe each event $80$ times
requires observing $\sim 44$ events per day, or one event every $\sim 30$ minutes.  The source
stars will be typically $I \lsim 17$, thus to achieve $2\%$ photometry requires 
observations of $\sim 10$ minutes on a $25\, {\rm{cm}}$ telescope.  
It is therefore possible to make measurements with 
$2\%$ photometry for all $\sim 100$ events.  As we discuss below, it is possible to measure
the projected speed for $\sim 70\%$ of these events.  Gould (1996) estimates that $\sim 15\%$ of
giant events could yield measurements of the proper motion, $\mu$.  Thus for a satellite
lifetime of $3$ years, we could expect to use combined parallax and proper motion information
to measure the mass and distance of $\sim 45$ lenses.

In our analysis, we choose to assume that the satellite in on an Earth-like 
orbit, separated from the Earth by a distance $\dsat$.
As we show in the Appendix, this orbit well approximates
a realistic orbit in which the satellite has a small velocity relative
to the Earth at the time of launch. 
Such an orbit is attractive on practical grounds since 
it requires the least amount of energy to attain.  Other orbits, such as
those inclined to the ecliptic plane, or with highly eccentric orbits, require a much
larger velocity relative to the Earth at the time of launch, 
and are therefore more expensive to achieve.  An Earth-like 
orbit is therefore optimal for the analysis because it is simplest to study and
is the most feasible.  As we demonstrate below, such an orbit is also 
advantageous because it allows for
parallax measurements towards the LMC and bulge equally well.

In a earlier paper, Boutreux \& Gould (1996) used Monte Carlo simulations of 
observations of microlensing events in the direction of the LMC to determine
the conditions needed to measure the projected speed, ${\tilde v}$ and the 
projected velocity, ${\bf{\tilde{v}}}$.  They found that with photometric 
precisions of $3\%$ for the Earth and $4\%$ for the satellite, and an
Earth-satellite separation of $> 0.5 {\rm{AU}}$ that it 
was possible to measure $\tilde v$ for $70\%$ and  ${\bf{\tilde{v}}}$ for
$50\%$ of LMC events.  They found that larger $\dsat$ were preferable, and that
parallax measurements were possible for a broad range of lens masses.

In this paper, we extend the Monte Carlo analysis of Boutreux \& Gould (1996) to 
observations of microlensing events in the
direction of the bulge in order to determine if similar satellite requirements
are needed for bulge and LMC observations.  
We have determined the conditions needed to break the
degeneracy for a majority of events, and thus to measure the projected speed,
$\tilde v$, and the projected velocity,
$\bf \tilde v$, to an accuracy of $\leq 10\%$.  We consider the effects
of the satellite requirements, the photometric precision and Earth-satellite separation,
as well as 
the mass of the lens and the position of the source star, in order to ensure that
the parallax measurements will be sensitive to a broad class of events.  We also
consider the time of year the observations are made, which, as we show, is a
critical parameter for bulge observations.  
As we discuss in \S\ 2.4, it is predominantly for this reason that the analysis 
of bulge observations is 
substantially more complicated than that of LMC observations.  
The reader may wish to review Boutreux \& Gould (1996) in order to become 
familiar with a similar problem in a simpler setting.

\chapter{Measuring $\bf \tilde v$}

The ability to measure $\bf \tilde v$ is dependent on a number of different 
parameters, several of which are tied together 
in intricate ways.  In this section, we 
define these parameters, and use a qualitative comparison between observations
towards LMC and observations towards the bulge to explain how these parameters
affect the measurement of the projected velocity.  In \S\ 3 we 
justify these ideas mathematically.
We quantify the ability to measure $\bf \tilde v$ by introducing the Degeneracy
Breaking Fraction (DBF).  The DBF can be thought of
as the percentage of events for
which the projected velocity can be measured to an accuracy of $\leq 10\%$
for a specified set of the independent parameters.

In order to measure the projected velocity, 
one must be able to break the degeneracy 
between the four possible solutions.  We formulate a constraint 
(derived from the difference in inverse timescales), and discard those
solutions which do not obey this constraint.  We require that 
those solutions allowed by the constraint have fractional errors
which are small, and
are close enough that their difference is unimportant.
If these requirements are met, we
consider the degeneracy broken.  We discuss the DBF
quantitatively in \S\ 2.3.

\section{Constraint Condition}

We distinguish the four discrete solutions by means of a constraint 
on the observables, $\Delta t$, $\Delta \beta$, and $\Delta \omega$.  Here
$\Delta \omega \equiv \omega' - \omega$ is the difference in inverse
timescales between the satellite and Earth.  Those solutions that are 
unallowed by the constraint at the $3 \sigma$ level are
distinguishable.  As we show below, the ability to break the degeneracy is
complicated by the fact that the constraint coefficients are 
strong functions of the time of year.

The constraint is (Gould 1995a)
$$
u_\Vert\omega \Delta t + u_{\bot} \Delta \beta - r \Delta \omega = 0,
\eqn\consta
$$
where ${\bf{u}}$ is the relative Earth-satellite velocity projected onto
the plane of the sky, $u_{\perp}$ and $u_{\parallel}$ 
are the components of ${\bf{u}}$ 
perpendicular and parallel to ${\bf r}$, and $r=|{\bf r}|$.  In order
to derive expressions for $u_{\perp}$, $u_{\parallel}$, and $r$,
we define $\alpha$ as the angle between the source vector ${\bf \hat s}$ and
the south ecliptic pole, and $\psi$ as the phase of the orbit such that
when $\psi = 0$, the projected displacement vector ${\bf r}$ is most closely
aligned with ${\bf \hat s}$.  We define $\Omega _{\oplus}=2\pi\,{\rm yr^{-1}}$,
and $R= |{\bf R}|$ as the magnitude of the Earth-satellite separation vector. 
The geometry is shown schematically in Figure \one.  
Using these definitions, we find,
$$\eqalign{
r &= R[1-\sin^2\alpha\cos^2 \psi]^{1/2},\cr
u_{\perp} &= -\Omega_{\oplus}R {R\over r}\cos\alpha,\cr
u_{\parallel} &= -\Omega_{\oplus}R {R\over r}\sin^2\alpha {\sin(2\psi)\over2}.}
\eqn\ru
$$  
Note that in  Gould (1995a) the sign of $u_{\parallel}$ was incorrectly given
as positive.  
We now recast the constraint equation in the form 
$$
\sum_{i=1}^{3}{\alpha_i a_i = 0},\eqn\ala
$$ 
where $\alpha_i$ are the coefficients of the constraint,
and $a_i$ are the event observables
$$
a_i=(\Delta t,\Delta \beta, \Delta \omega).\eqn\obsers
$$
Using this, we find the simpliest
form for the constraint coefficients is,
$$
\alpha_i =
[\Omega_{\oplus}\omega \sin^2\alpha {\sin(2\psi)\over2} ,\,\,
\Omega_{\oplus}\cos\alpha ,\,\,(1-\sin^2\alpha\cos^2 \psi) ].
\eqn\coeff
$$

\topinsert
\mongofigure{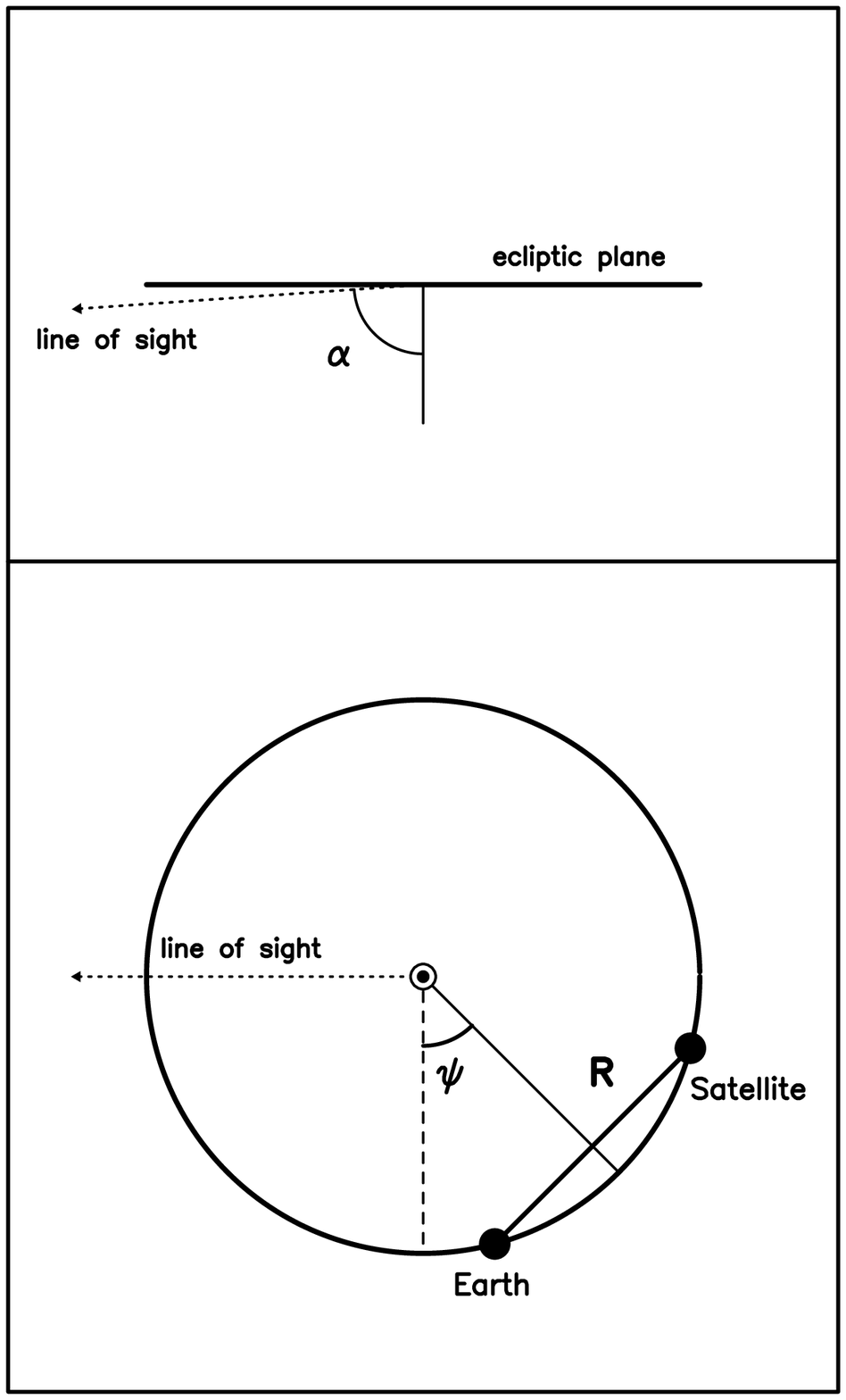}{6.4}{7.1}{6.4}
{The geometry of the orbits of the Earth and the satellite.  Note in 
particular the angle $\psi$, defined as the angle such that 
when $\psi = 0$, the projected displacement vector ${\bf{r}}$ is 
most closely aligned with the line of sight.
}
\endinsert

In order to break the degeneracy, the constraint equation {\ala} must
distinguish between the four degenerate solutions at the $3\sigma$ level.
But, as we have shown above, the constraint coefficients $\alpha_i$ are
dependent on $\alpha$ and $\psi$, the line of sight of observations and the
phase of the orbit.  Furthermore, from equations  \delx\ and {\ru}, 
we see that the vector displacement $\Delta {\bf x}$ is a
function of $\alpha$ and $\psi$.  Thus, because
${\Delta\bf{x}}  \equiv (\omega \Delta t,\Delta \beta)$, the event observables
$a_i$ are also functions of $\alpha$ and $\psi$.  All of these contributions
are obviously difficult to disentangle, and thus it is not a straightforward
matter to describe the conditions under which the constraint equation
\ala\ can break
the degeneracy.

\section{Fractional Error}

We define the fractional difference as the
difference between the allowed solution and the true solution, divided 
by the true solution. In order that the degeneracy be broken, 
we require that the fractional
errors in the allowed solutions be $< 10\%$.   
The behavior of the fractional errors is easier to understand
than the behavior of the constraint.
Roughly speaking, as $\Delta x$ decreases, 
the fractional errors rise.  Thus, from equation \delx\ we can see that as 
$r$ decreases, or as $\tilde \re$ increases, the 
fractional errors rise.  From equation {\ru}, we see that the 
projected Earth-satellite separation, $r$, is a 
function of direction of the source, $\alpha$, and the phase of the
orbit, $\psi$.  Therefore these parameters affect the fractional error, 
and thereby affect the DBF.   

\section{DBF and Errors}

We define the DBF as the fraction of events for which the degeneracy can
be broken for a specified set of parameters.  We consider that the degeneracy 
is broken if 
the difference between all allowed 
solutions and the true solution is 
less than $20\%$ of the true solution, and
the fractional error in $\Delta x$ of each allowed
solution is less than $10\%$.  In principle, it could be difficult to 
characterize the errors in the DBF precisely.  This is because there are
two distinct types of errors that must be considered.  The first is an
intrinsic error that arises from the uncertainty in the values of each 
distinct solution.  This error is approximately
Gaussian distributed.  The second error
is characterized by two discrete allowed
solutions, and hence is highly non-Gaussian.
Taken together, these errors can not be described simply.

We would like to be able to characterize our errors as $10 \%$ at
the $1 \sigma$ level.  We then consider all possible events with $10 \%$ 
fractional error but with multiple allowed solutions.  If two allowed 
solutions 
are separated by $20 \%$, then the $1 \sigma$ error contour is strongly 
perturbed by the presence of a second solution, while the $3\sigma$ contour
is barely affected.  On the other hand, if the discrete
solutions are $5\%$ apart,
then even the $1 \sigma$ contours are essentially
unaffected.  Thus, at first sight,
the threshold for the allowed separations of distinct solutions appears
to depend sensitively on the error contour of interest.  In practice, however,
the DBF is very nearly the same for $5\%$ and $20\%$ thresholds.  
See Figure \two, below.
In this paper we adopt $20\%$, which corresponds roughly to focusing on
the error contours at the $2\sigma$ level.

\section{LMC vs. Bulge}

Observations towards the Galactic bulge are markedly different from 
those towards the LMC.  For observations towards the bulge,
the results are complicated by the fact that
the ability to break the degeneracy is a strong function of the time of year;
this is not the case for observations towards the LMC.

For observations towards the LMC, $\alpha \sim 0$, and from 
equation {\ru}, we find 
$r = R$, $u_\Vert = -R\Omega_{\oplus} $, and $u_{\bot}=0$.
These expressions are obviously independent of $\psi$.
Using equation {\coeff}, the constraint coefficients become, 
$\alpha_i=(0,\Omega_{\oplus},1)$. 
Thus the ability to distinguish between the degenerate
solutions is independent of the phase of the orbit, 
and depends only on the photometric
precision, the number of observations per crossing time, and the 
Earth-satellite separation.  Given a certain set of parameters, it is 
therefore possible to quantify the fraction of events for which the degeneracy
can be broken, regardless of the phase of the orbit.  
Furthermore, because $r$ does not vary with $\psi$, 
for any given event the fractional errors are also constant
throughout the orbit.

The situation is considerably more complicated for bulge observations.  For
this case, $\alpha \sim 90^\circ$.  From equation \ru, we find that the 
values of $r$ and $u_{\bot}$ vary as a function of $\psi$.  Therefore
the coefficients of the 
constraint equation {\coeff}, and hence the ability of the constraint to distinguish 
between the solutions, are a strong function of the phase of the orbit.
For example, consider Baade's window, for which $\alpha = 84^{\circ}$.   
For $\psi = 0$ or $180^{\circ}$, 
we can approximate the constraint equation as
$$
\Delta \beta \simeq - {{\Delta \omega}\over{10\Omega_{\oplus}}},
\eqn\constzero
$$
whereas for $\psi = 90^{\circ}$ or $270^{\circ}$,
$$
\Delta \beta \simeq - {{10 \Delta \omega}\over{\Omega_{\oplus}}}.
\eqn\constnin
$$
We can see from these relations the effect of $\psi$ on the ability of the
constraint to distinguish between the degenerate values of $\Delta \beta$. 
For a specified uncertainty in the value of $\Delta \omega$, the uncertainty
in $\Delta \beta$ is 100 times less when $\psi = 0$ or $180^{\circ}$, than
when $\psi = 90^{\circ}$ or $270^{\circ}$.  Thus we can expect that, 
in general, it is easier to distinguish
between the degenerate solutions when $\psi = 0$ or $180^{\circ}$ than when
$\psi = 90^{\circ}$ or $270^{\circ}$.  For intermediate values of $\psi$, 
we can interpolate the behavior of the constraint accordingly.

For bulge observations, the intrinsic errors are also 
strongly dependent on $\psi$.  From equation \ru\ we see that, for 
observations near $\alpha = 90^{\circ}$, the projected
Earth-satellite separation, $r$, varies approximately as $\sin {\psi}$.  
Thus, from equation {\delx}, for $\psi \sim 0$ or $180^{\circ}$, 
the value of $\Delta x$ is comparatively small, and therefore
the fractional errors are comparatively large, 
whereas for $\psi \sim 90^{\circ}$ or 
$270^{\circ}$, the value of $\Delta x$ is comparatively large, and thus
the fractional errors are comparatively small.  As before, we can interpolate
the behavior of the fractional errors for intermediate values of $\psi$.

The effects of $\psi$ on the constraint condition and fractional errors are
in competition with each other.  We therefore expect the DBF, which
is a combination of both of these considerations, to have four minima: when 
the fractional errors dominate, at $\psi = 0$ or $180^{\circ}$, {\it{and}}
when the constraint on $\Delta \beta$ is poor, at 
$\psi \simeq 90^{\circ}$ or $270^{\circ}$.   We expect the maximum values
of the DBF to occur where the fractional errors are not dominant, yet the 
constraint still distinguishes between the degenerate solutions reasonably 
well.  In the next section, we review these qualitative arguments in a 
more precise form.

\chapter{Error Analysis}

In this section we give an overview of the mathematical analysis
underlying the Monte Carlo code.  For a more complete description, we
refer the reader to \S\ 2 of
Boutreux \& Gould (1996); much of the
analysis given therein is entirely applicable to the present considerations.
The reader who is primarily interested in the results can skip this section 
entirely.

The flux $F(t)$ from a microlensed star is a function of five parameters,
$a_i=(t_0,\beta,\omega,F_0,B)$, with the form,
$$
F(t;t_0,\beta,\omega,F_0,B)=F_0A[x(t;t_0,\beta,\omega)]+B,\eqn\flux
$$
where $F_0$ is the unlensed flux, and $B$ is the light from any additional
unlensed sources that are not resolved.  The flux from 
the satellite can also be described in this manner, by five additional
parameters $a_i'=(t_0',\beta',\omega',F_0',B')$. Using measurements simulated by
our Monte Carlo method, we then fit the parameters by minimizing 
$\chi^2$.  By differentiating $\chi^2$ in the neighborhood of the solution, 
one can obtain $c_{ij}$, the covariance matrix of the $a_i$ for both the 
Earth and the satellite.

We assume that the Earth and satellite have the same filters, implying that
$F_0=F_0'$ and $B=B'$.  Each of these constraints can be written
${\sum_{i=1}^{10}\alpha_i a_i = 0}$,  where $a_i$ are now the ten parameters for the Earth and
the satellite, and $\alpha_i$ are the constraint coefficients.  The constraint
coefficients for $F_0=F_0'$ are 
$\alpha_i = (0,0,0,1,0,0,0,0,-1,0)$.  We impose these constraints 
on the $10 \times 10$ covariance matrix
$c_{ij}$, forming a new matrix $\tilde c_{ij}$,
$$
\tilde c_{ij} = c_{ij} - 
{{\sum_{l=1,k}^{10}{{c_{il}\alpha_l c_{jk} \alpha_k}}}
 \over
{\sum_{m,n=1}^{10}{{c_{mn} \alpha_{m} \alpha_{n}}}}}. \eqn\ctilda
$$

Neglecting the relative Earth-satellite motion, we find that 
there is a two-fold degeneracy 
in the magnitude $|\Delta\beta|=\Delta \beta_{\pm}$, and another two-fold
degeneracy in the sign.  Thus there are four possible solutions, 
$\pm\Delta\beta_{\pm}$.  In our simulations, we know the true solution, which
for the illustrations given here we take to be $+\Delta \beta_-$ or ($-+$) for
short.  However, 
we initially assume that only the magnitudes of $\beta$ and $\beta'$ are known,
thus making all four possible values of $\Delta\beta$ equally probable.  We use
the constrained covariance matrix, $\tilde c_{ij}$, to form a new covariance
matrix of the quantities $a_i=(\Delta t, \Delta \beta, \Delta \omega)$; we
call this new matrix $c_{ij}^{-+}$. Due to the fact that there are four 
possible values of $\Delta \beta$, we form four different matrices
$c_{ij}^{\pm \pm}$.  We
then test to see if the solutions are distinguishable from the true solution,
using the constraint imposed by the relative motion of the Earth and the 
satellite, equations {\ala}\ --\ {\coeff}.

The general form of $\chi^2$ can be written as
$$
\chi^2_{-+}={(\alpha_2\delta\beta^{-+})^2\over
{\sum_{m,n=1}^{3}{c^{-+}_{mn}\alpha_m\alpha_n}}},
\eqn\chigen
$$
where the $\alpha_i$'s are the coefficients from the constraint condition, 
and $\delta\beta^{-+}$ is the difference between the true and trial values
of $\Delta\beta$.  Because $\chi^2$ is a strong function of the phase of the 
orbit, it is not possible to give $\chi^2$ in a simple form. 
It is illuminating, however, to examine the form of
$\chi^2$ at the values of $\psi$ for which it is at maximum and minimum.  
 For $\psi=({90^{\circ}},{270^{\circ}})$,  the constraint coefficients are: 
$$
\alpha_i = (0,\Omega_{\oplus}\cos\alpha,1)\eqn\alpa
$$
and for $\psi=(0,180^{\circ})$,
$$
\alpha_i = (0,\Omega_{\oplus},\cos\alpha).\eqn\alpb
$$
We define $q = (\sigma_{\omega}/\sigma_{\beta}t_e{\Omega_{\oplus}})$, 
where $\sigma_{\omega} = c^{1/2}_{33}t_e$, and $\sigma_{\beta}=c^{1/2}_{22}$.  
Using these expressions, and assuming (as is generally the case),
that the off-diagonal 
covariances ($c_{ij},i\not =j$) are small, 
the equation for $\chi^2$ can be written for
each case as,

$$\eqalign{
\chi^2_{min} =\chi^2(90^{\circ},270^{\circ})
&={ { \delta \beta / \sigma_{\beta}} \over { 1 + ( q / \cos{\alpha} )^2}}   
\cr
\chi^2_{max}=\chi^2(0,180^{\circ})
&={ { \delta \beta / \sigma_{\beta}} \over { 1 + ( q \cos{\alpha} )^2}}}
\eqn\chimm
$$
For typical events, $q \sim 6$ for the
$\pm\Delta\beta_-$ solutions; this value is roughly
independent of $\psi$.

Note that $\chi^2_{min}/\chi^2_{max} \sim (\cos{\alpha}/q)^2$ 
(for typical time 
scales).  Thus for 
observations toward the bulge, $\alpha \simeq 84^\circ$,
$\chi^2$ can vary by a factor of $10^2 - 10^4$ over
the course of the orbit.

We calculate the $\chi^2$ for all four solutions, and consider only those 
solutions for which $\chi^2 \leq 9$.  We then calculate the intrinsic error
for these allowed solutions.  The fractional error in the scalar solution 
can be approximated by $\langle(\delta\Delta x)^2\rangle^{1/2}/\Delta x$, where
again $\Delta{\bf{x}} \equiv (\omega\Delta t, \Delta \beta)=(a_1,a_2)$, and
$\Delta{x}=|\Delta{\bf{x}}|$.  Using the analog of equation \ctilda,
$$
\tilde c_{ij}^{-+} = c_{ij}^{-+} - 
{{\sum_{l=1,k}^{3}{{c_{il}\alpha_l c_{jk} \alpha_k}}}
 \over
{\sum_{m,n=1}^{3}{{c_{mn} \alpha_{m} \alpha_{n}}}}}, \eqn\ctildatwo
$$   
we apply
the constraint \ala\ to $c^{-+}$ to form $\tilde c^{-+}$, where
$\alpha_i$ is now given by equation \coeff, and $a_i = (\Delta t, \Delta \beta, \Delta
\omega)$.
In terms of $\tilde c_{ij}^{-+}$, the fractional error 
becomes
$$
{{\langle(\delta\Delta x)^2\rangle^{1/2}}\over{\Delta x}}=
{{(\sum\nolimits_{i,j=1}^{2} \tilde c_{ij}^{-+} a_i a_j)^{1/2}}\over{(a_1)^2 +
(a_2)^2}}.\eqn\scerr
$$
From this we can see that as the quantities $(a_1,a_2)$ become small, the 
fractional error in the solutions rise.  Recall that for $\psi \sim 0$, the
projected Earth-satellite separation is small (relative to the true 
separation), and therefore $\Delta x$ is small.  Thus the intrinsic errors
in the solutions become more important near $\psi = 0$.

We approximate the fractional error in the vector solution 
by 
$\langle(\delta\Delta {\bf{x}})^2\rangle^{1/2}/\Delta x$.  
This can be written
as,
$$
{{\langle(\delta\Delta {\bf{x}})^2\rangle^{1/2}}\over{\Delta x}}=
 \left[{{\tilde c_{11}^{-+}} + 2{\tilde c_{12}^{-+}} +  
{\tilde c_{22}^{-+}}} \over {(a_1)^2 +(a_2)^2}\right].
\eqn\veerr
$$

\chapter{Bulge and Disk Models}

In order to calculate the DBF using the Monte Carlo 
simulation, we adopt appropriate models for the velocity and 
density distributions of the disk and bulge of the Milky Way.  

For the density distributions, we choose
a Bahcall model for the disk.  For the bulge, we 
adopt a `revised' COBE model.  This model uses a barred
bulge density distribution for the outer parts of the bulge 
($r > 0.7 \, {\rm{kpc}}$)
and a Kent model for the inner part of the bulge.  For a more detailed 
description, see Table 3 from Han \& Gould (1995).  

We assume the velocity distributions to be Gaussian, with mean velocity 
$\bar v$, and velocity dispersion $\sigma $.
For the disk, we use $(\bar{v}_x,\bar{v}_y,\bar{v}_x)=(0,220,0)$; 
with dispersions of
$(\sigma_x,\sigma_y,\sigma_z) =(40,30,20) \kms$.  For the bulge,
$(\bar{v}_x,\bar{v}_y,\bar{v}_x)=(0,0,0)$, while $(\sigma_x,\sigma_y,\sigma_z)
=(110,82.5,66.3) \kms$.
The motion of the Sun is also taken into account, and is 
$(v_x,v_y,v_z)=(9,231,16)\kms$ (Mihalas \& Binney 1981).

There are two distinct types of events  
that are seen towards the bulge.  These 
are bulge-bulge events, where a bulge star is being lensed by bulge lens, and
disk-bulge events, where a bulge star is being lensed by a disk lens.
These two situations produce dissimilar events, and we therefore 
consider them separately.

For disk-bulge events, the position of the source star, $\dos$ is confined
to a relatively narrow range as compared to the range of possible values
for $\dol$.  Thus we can approximate the location of the
source stars as being fixed at the 
galactocentric radius, $R_{0} = 8\,\rm{kpc}$.  The frequency of lensing events
scales as
$\rho(r_{\rm{l}})\re v \propto \rho(r_{\rm{l}}){[z(1-z)]^{1/2}}\,v$, where
$v$ is the transverse velocity of the lens, $r_{\rm{l}}$ is the distance of 
the lens from the galactic center (for lines of sight near the galactic center,
$r_{\rm{l}} \simeq \dls$), and $\rho(r_{\rm{l}})$ is the mass
density at $r_{\rm{l}}$.
Thus to simulate an event, we draw the distance 
to the lens randomly from the probability
distribution $\rho(r_{\rm{l}}){[z(1-z)]}^{1/2}$.  The
velocities of the lens and the source are then drawn randomly according to 
the parameters described above, and the transverse velocity calculated.  
We weight the DBF by the transverse speed in order to include its effect on the
event rate.

For bulge-bulge events, the frequency of lensing events depends on the
mass density at the position of the source, the mass density at the position
of the lens, the size of the Einstein ring, 
and the transverse velocity of the lens.  In addition, as the
distance from the observer increases, the volume element along the line
of sight also increases, and therefore so does the total number of stars. 
Thus the event
rate scales as $\rho(r_s)\rho(r_l) \re \dos^2 v$.  Here $r_s$ is the
distance of the source from the galactic center, and $\dos^2$ provides the
contribution due to the increasing volume element.  Thus, the source-lens 
pair position ($\dos,\dol$) is drawn randomly 
from the distribution $\rho(r_s)\rho(r_l) {[z(1-z)]^{1/2}} \dos^2 $.  
The resulting DBF is weighted by the transverse velocity, as above.

\chapter{Monte Carlo}

For the simulation, we assume that the observations begin when the source 
enters the Einstein ring radius as seen from the Earth.  The observations 
continue until the source-lens separation is $\geq 3\re$ as seen from both the
Earth and the satellite.  The impact parameter, $\beta$, is either fixed, or
chosen randomly in the range $ 0 < \beta < 0.7$.  This range of $\beta$ is 
consistent with that of observed microlensing events. Those events for 
which the time of maximum magnification for the satellite has already passed
when the observations begin, that is, $(1 - \beta^2)^{1/2} < \omega \Delta t$,
are considered unresolvable.  Those events for which $\beta' >1.5$ are also
considered unresolvable.

The four possible solutions are considered separately, and the $\chi^2$ 
calculated for each.  Those solutions for which $\chi^2 >9$ are 
distinguishable from the true solution, and are therefore unallowed.  
We then determine if the allowed solutions are sufficiently similar that
it is unimportant to break the degeneracy, and if the 
fractional errors in the solutions are intolerably large.  The scalar
difference between the solutions is given by,
$$
\delta(\Delta x) =  { {|\Delta x|}_{\rm{allowed}} - {|\Delta x|}_{\rm{true}} },
 \eqn\scadif
$$
whereas the vector difference is given by,
$$
\delta(\Delta {\bf{x}}) =  | {\Delta{\bf{x}}}_{\rm{allowed}} - 
{\Delta {\bf{x}}_{\rm{true}}}|. 
\eqn\vecdif
$$   
If the difference between all allowed 
solutions and the true solution is less than $20\%$ of the true solution, and
if the fractional error in $\Delta x$ of each allowed
solution is less than $10\%$,
we consider that the degeneracy is broken.
For each set of parameters, the Monte
Carlo simulation was run 5000 times, resulting in errors of $< 1\%$.

\chapter{Observational Considerations}

Because the bulge is located at ecliptic angle 
$\alpha \sim 90^{\circ}$, there will be times of the year that observations
will be impossible, due to the position of the Sun.  To quantify
this effect, we define an allowed range of the ecliptic longitude, $\Lambda$,
for the Earth and the satellite.  
When the position of the Earth or the satellite is
within this range, observations are considered impossible.  The full 
width of this
range is chosen to be $90^{\circ}$ (3 months) for the Earth, and $60^{\circ}$
(2 months) for the satellite.
In our simulations, we consider lines of sight towards the bulge, 
for which $\Lambda = 270^{\circ}$.  The forbidden ranges for
the Earth and satellite are then,
$$\eqalign{
90^{\circ}-45^{\circ} &\leq \Lambda_{E} \leq 90^{\circ}+45^{\circ}\cr
90^{\circ}-30^{\circ} &\leq \Lambda_{S} \leq 90^{\circ}+30^{\circ}.}\eqn\rest 
$$
Note that $\Lambda = 0$ is the vernal equinox.  The time of the year when
either the Earth or the satellite is within the forbidden range is dependent
on the Earth-satellite separation, $\dsat$.  For $\dsat \leq 75^{\circ}$,
the forbidden ranges of the Earth and satellite overlap. Conversely, 
for $\dsat \geq 75^{\circ}$, the forbidden ranges 
are separated by a gap of $\Delta \Lambda = \dsat - 75^{\circ}$.  Thus values
of $\dsat \geq 75^{\circ}$ are somewhat less desirable due to the fact that
the times of year when observations are possible are separated into two 
ranges, thereby increasing the probability that the observations of an 
event will be foreshortened due to the Earth or the satellite moving behind the
Sun.  We will consider this issue more fully
in \S\ 7.1, when we discuss the optimal 
Earth-satellite separation.

\chapter{Results}

There are eight parameters that affect the degeneracy breaking fraction; 
$\alpha$, the angle between the source and the south ecliptic pole; $\psi$, 
which is related to the phase of the orbit; $\sigma_E$ and $\sigma_S$, 
the photometric precisions of the Earth and satellite observations, 
respectively; $\beta$, the impact parameter as seen from the Earth; 
$d_{sat}$, the Earth-satellite separation distance; $M$, 
the mass of the lens; $N_{meas}$, the number of measurements 
per Einstein ring crossing time.

For all of the parameters (neglecting $\psi$ for the moment), 
we choose typical values that we
consider to be realistic.  These values are used in all simulations unless
varied explicitly.  A value of $N_{meas} = 20$ corresponds to
frequency of measurements in the range of 0.5 to 4.0 days for typical 
events.  We choose $(\sigma_E,\sigma_S)=(1\%,2\%)$ as our 
photometric precision.  The photometric precision obviously depends on 
the magnitude of the source star; these are typical values of photometric
precision for giant sources.  
We choose 
$M=0.3M_{\odot}$ as a representative mass of lenses for observations towards
the bulge.  The default Earth-satellite separation is $d_{sat}= 60^{\circ}$ 
(2 months).
The maximum impact parameter is set at $\beta_{max}=0.7$.  For
Baade's window, $\alpha=84^{\circ}$.

Observations towards the bulge are unique in that they are a strong function
of the time of year, as discussed above.  Due to this 
intrinsic dependence of the DBF on $\psi$, we include
$\psi$ as a free parameter in all further calculations.

In order to elucidate the effect of the time of the year on the 
DBF, we present in Figure {\two} the DBF against
$\psi$ for our fiducial parameters.  Included in Figure {\two} are 
the results for both the disk-bulge and bulge-bulge cases.  
Also included is the DBF against
$\psi$ for disk-bulge events assuming a fractional difference threshold of $5\%$ (see \S\ 2.3).
Within the limits of the Monte Carlo 
errors, we find that in the range $0 \leq \psi \leq 180^{\circ}$ the curve is 
equivalent to that for the range 
$180^{\circ} \leq \psi \leq 360^{\circ}$.  We utilize 
this symmetry by restricting the range of $\psi$ in further calculations to
$0 \leq \psi \leq 180^{\circ}$.  There is, however, a slight asymmetry 
around $\psi = 90^{\circ}$ (and similarly around $\psi = 270^{\circ}$). 
For the speed degeneracy, this asymmetry makes no
significant contribution, and we disregard it.  We will consider this
effect more fully in \S\ 8, where we discuss the velocity degeneracy.
For $\psi \sim 0, 180^{\circ}$, we see from equation \ru\ that the 
projected Earth-satellite velocity is at a maximum, and as a result
the constraint given in equation \consta\ is very effective in distinguishing 
between the possible solutions.  The 
projected Earth-satellite separation is at a minimum, however, 
driving up the fractional
errors in the solutions.  Thus near $\psi \sim 0, 180^{\circ}$, the DBF is 
dominated by the intrinsic errors.  As $\psi$ increases, the projected
Earth-satellite separation increases, and thus the intrinsic errors 
decrease.  The projected Earth-satellite velocity decreases as $\psi$ 
increases, and as a result the constraint becomes less effective in 
distinguishing between the degenerate solutions.  For 
$\psi \simeq 20^{\circ}$ the DBF is at a maximum; the constraint still
distinguishes between the solutions for a majority of the events, while the 
intrinsic errors are typically $\leq 10\%$.  
When $\psi = 90^{\circ}$, the projected
Earth-satellite separation is equal to the true Earth-satellite separation,
and the intrinsic errors are typically $ \leq 5\%$.  The projected 
Earth-satellite velocity is at a minimum at $\psi = 90^{\circ}$, and the
constraint distinguishes poorly among the solutions (as compared to
$\psi \sim 0$).  Thus, at 
$\psi = 90^{\circ}$, the DBF is dominated by the difficulty in distinguishing
 between
the individual solutions.  The analysis of the 
DBF curve in the range
$0 \leq \psi \leq 90^{\circ}$ is identical to that for the rest of
the orbit.  

The analysis of bulge-bulge events is exactly the same as for 
disk-bulge events.
We can understand how disk-bulge and
bulge-bulge events differ by considering the typical size of the projected
Einstein ring in each case.  For disk-bulge events, the median value for
the distance to lens is $\dol \sim 5 \,{\rm {kpc}}$, and the distance to the 
source is $\dos = 8\, {\rm {kpc}}$, 
thus, from equations \einrad\ and {\redein}, we
find that, for disk lenses,
${\tilde \re} \simeq 6.0 {\rm{AU}}(M/0.3M_{\odot})^{1/2}$.  For bulge-bulge 
events, $\dol \sim 7\,{\rm {kpc}}$, $\dos \sim 9\,{\rm {kpc}}$, so that
${\tilde \re} \simeq 9.0 {\rm{AU}}(M/0.3M_{\odot})^{1/2}$.  Therefore the 
typical size of the projected Einstein rings will be larger for bulge-bulge
events by a factor of $\sim 1.5$.  This will affect the value of the DBF
in two distinct ways.  For phases of the orbit such that $\psi \sim 0$, 
where intrinsic errors dominate, the DBF for bulge-bulge events will be 
smaller than that for disk-bulge events, due to the larger
typical values of $\tilde \re$ (and therefore larger fractional errors).  For
phases of the orbit such that $\psi \sim 90^{\circ}$, we expect that the
DBF for bulge-bulge events will be, in general, higher than the corresponding DBF
for disk-bulge events, due to the fact that the degenerate $\Delta \beta$ solutions
are better separated in the case of the bulge-bulge events
than those for disk-bulge events. 

For disk-bulge events,
we find that for our fiducial parameters, 
it is possible to resolve $> 87\%$ of
the events during most times of the orbit, and $> 70\%$ of the events 
over the entire orbit.  For bulge-bulge events, we find values of $> 90\%$
for most of the orbit and $> 60\%$ for the entire orbit.

\topinsert
\mongofigure{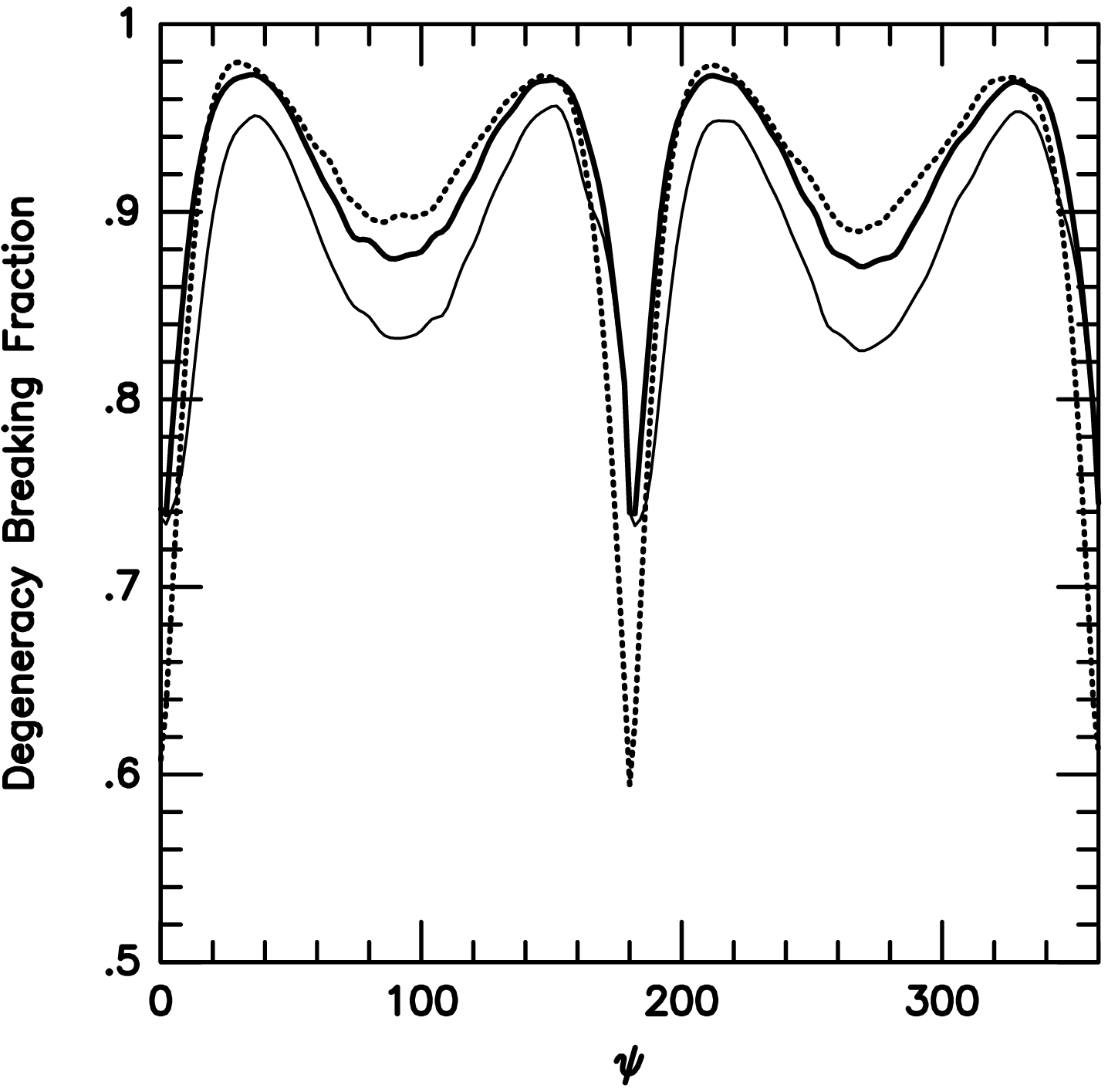}{6.4}{6.0}{6.4}
{
Degeneracy Breaking Fraction (DBF) as a function of $\psi$ for
disk-bulge events (solid bold curve) and bulge-bulge events (dashed curve).  All other
parameters are fixed at $n_{obs} = 20$, $(\sigma_E,\sigma_S)=(1\%,2\%)$, 
$M=0.3M_{\odot}$, $d_{sat}=60^{\circ}$, $\beta_{max}=0.7$, and
$\alpha=84^{\circ}$.  The DBF is the fraction of events for which
the allowed solutions differ by $< 20 \%$ and each of these has intrinsic
error $< 10 \%$.  The lower solid curve is the DBF as a function of
$\psi$ when allowed solutions differ by $< 5\%$, and each has intrinsic
error $< 10 \%$.
}
\endinsert

Figure {\three} represents the DBF as a function of $\psi$ and $\beta$ for disk-bulge events.  
For this 
figure, $\beta$ is not randomly chosen, but rather held fixed for each point
in the curve.  Contrary to the results for the LMC (Boutreux \& Gould 1996),
where it was found that
events with $\beta \geq 0.7$ are much harder to resolve, there is no one 
value of $\beta$ for which the DBF drops below a certain value.  In fact, 
for $\psi = 90^{\circ}$, even those events for which $\beta \sim 0.9$
are resolvable 80\% of the time, whereas less than 20\% of these events
are resolvable when $\psi = 10^{\circ}$.  In other simulations, we choose 
$\beta$ randomly between $0 \leq \beta \leq 0.7$ because the majority of
events seen so far have values of $\beta$ that are in this range, but we 
emphasize that, under the proper conditions, even those events 
that have $\beta > 0.7$ may still be resolvable.

\topinsert
\mongofigure{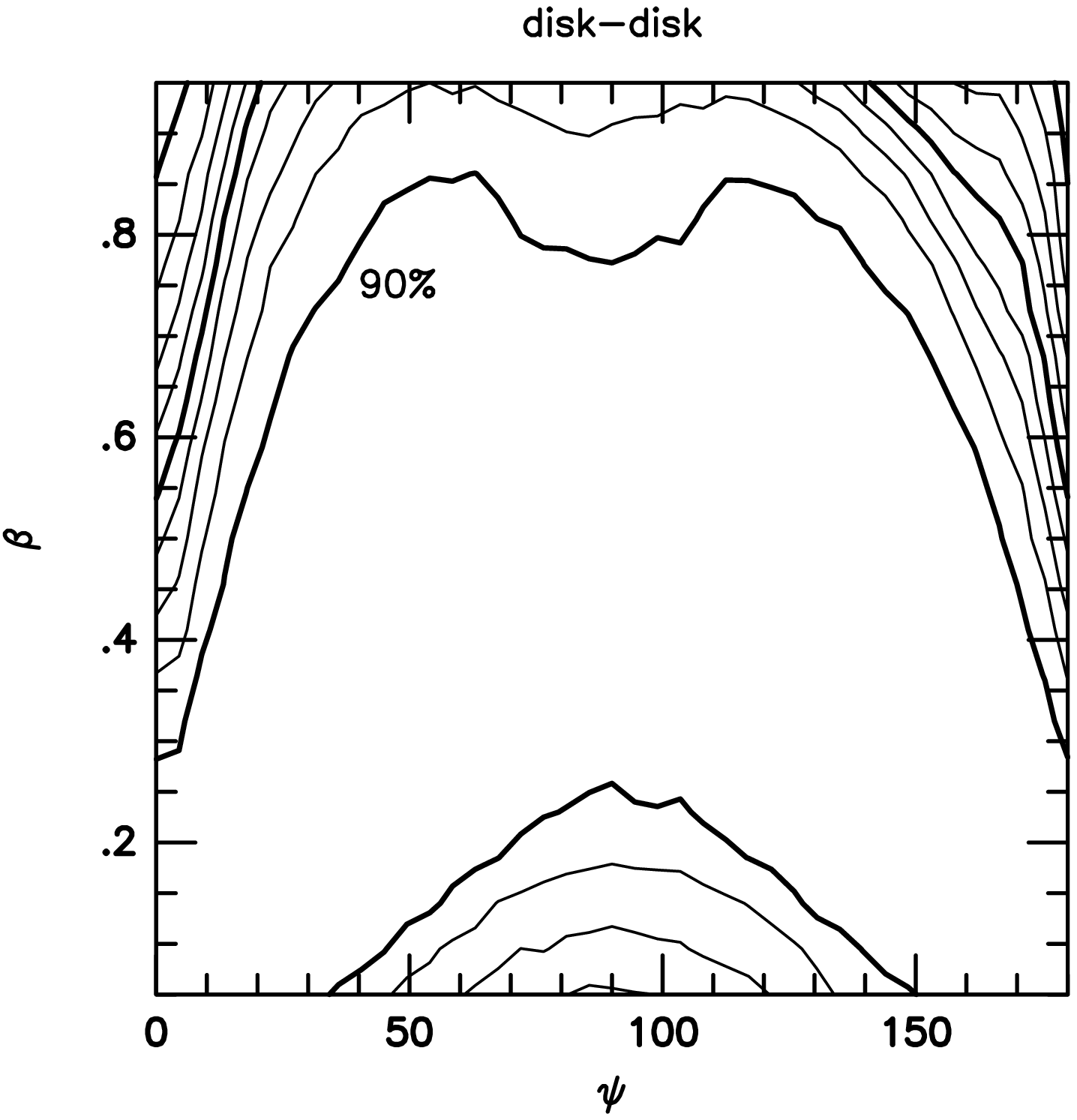}{6.4}{6.0}{6.4}
{
DBF as a function of $\psi$ and
$\beta$ for disk-bulge events. 
The labeled bold contour represents a DBF of 90\%.  Other bold contours
represent DBFs of 50\% and 10\%.  Intermediate contours are spaced at
intervals of 10\%.  All other parameters are held fixed at our fiducial
parameters, as in Fig.\ {\two}.
}
\endinsert

Figure {\four} represents the DBF as a function of $\psi$ and $M$ for disk-bulge events.
For the mass, we focus on the range $0.01M_{\odot} -1M_{\odot}$; 
this corresponds
to the best estimate of the range of masses for events already observed.
For our fiducial value, $M=0.3 M_{\odot}$, the DBF is $\sim 90\%$ for
the majority of the orbit. For disk-bulge events, the projected Einstein
ring radius is ${\tilde \re} \simeq 6.0 {\rm{AU}}(M/0.3M_{\odot})^{1/2}$.   
Consider first phases of the orbit for which
$\psi \sim 0$, where the fractional errors dominate the
behavior of the DBF.  For these values of $\psi$ (and $R \sim 1 {\rm{AU}}$),
the projected Earth-satellite separation is 
$r \sim 0.1{\rm{AU}}$.  The fractional errors become important when
$\Delta x \ll 1$.  A quick calculation shows that, 
for masses greater than $M \geq 0.1M_{\odot}$, the magnitude of
the vector displacement is $\Delta x = r/{\tilde \re} < 0.1$, and thus
the fractional errors begin to dominate, thereby lowering the DBF.
For masses $M \leq 0.03M_{\odot}$, $\Delta x = r/{\tilde \re} \sim 1$, and
thus the fractional errors are unimportant, and the DBF is $> 90 \%$.
Consider now phases of the orbit for which $\psi \sim 90^{\circ}$.  
For the range of masses considered here, $\Delta x \geq 1$, and thus the 
fractional errors are never large.  Considering increasingly smaller masses,
we find that more of the events are unresolvable, due to the fact that 
the projected Earth-satellite separation, $r$, becomes increasingly larger
than the Einstein ring radius, ${\tilde \re}$, and the event as
seen from the satellite will more frequently fall outside of the Einstein ring.
Thus the DBF drops rapidly from $90\%$ to $40\%$ from $M=0.3M_{\odot}$ to
$M=0.03M_{\odot}$.     

\topinsert
\mongofigure{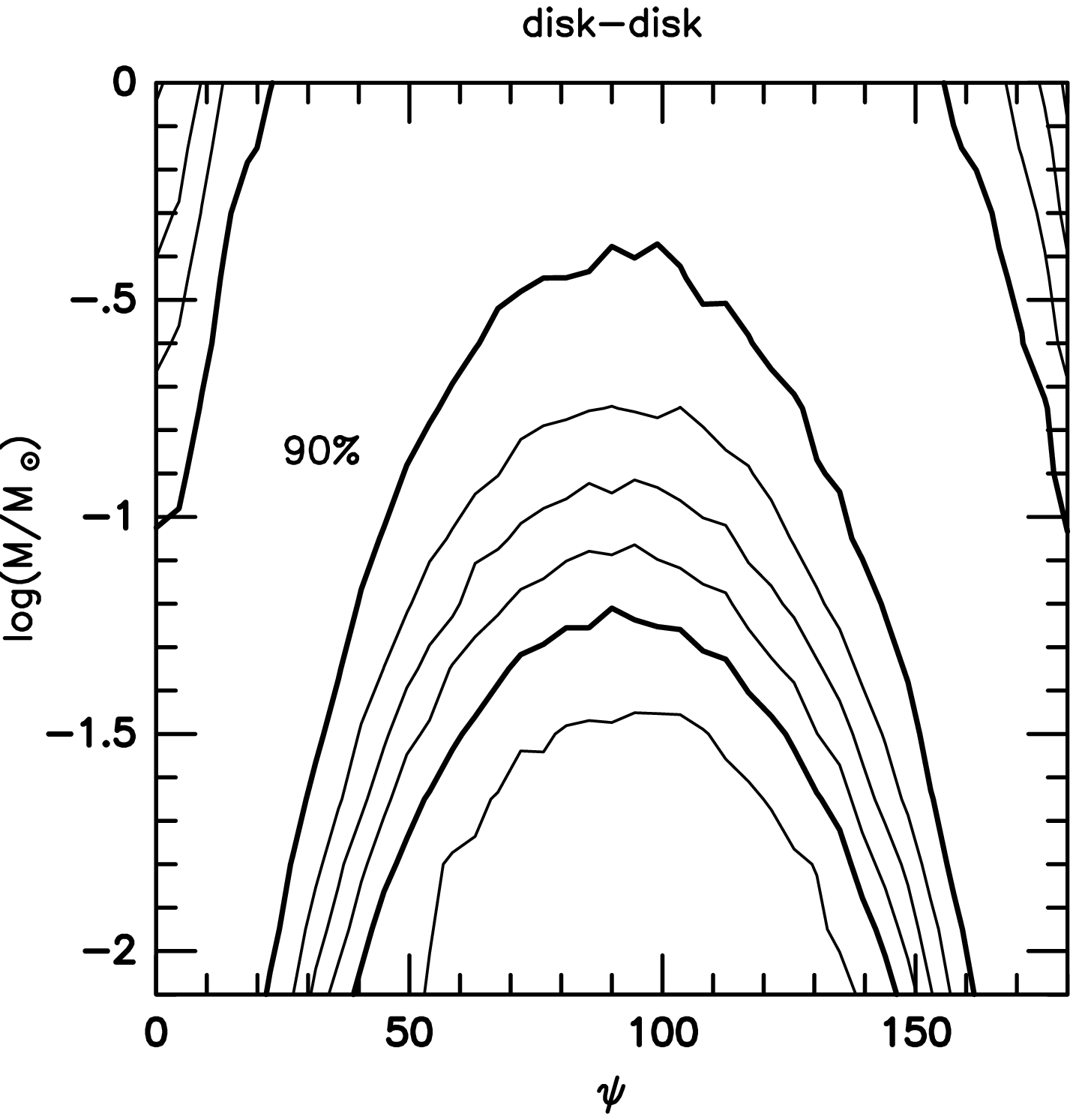}{6.4}{6.0}{6.4}
{
DBF as a function of $\psi$ and
$\log{M/M_{\odot}}$ for disk-bulge events. 
Contours are the same as those in Fig.\ {\three}. 
}
\endinsert

\topinsert
\mongofigure{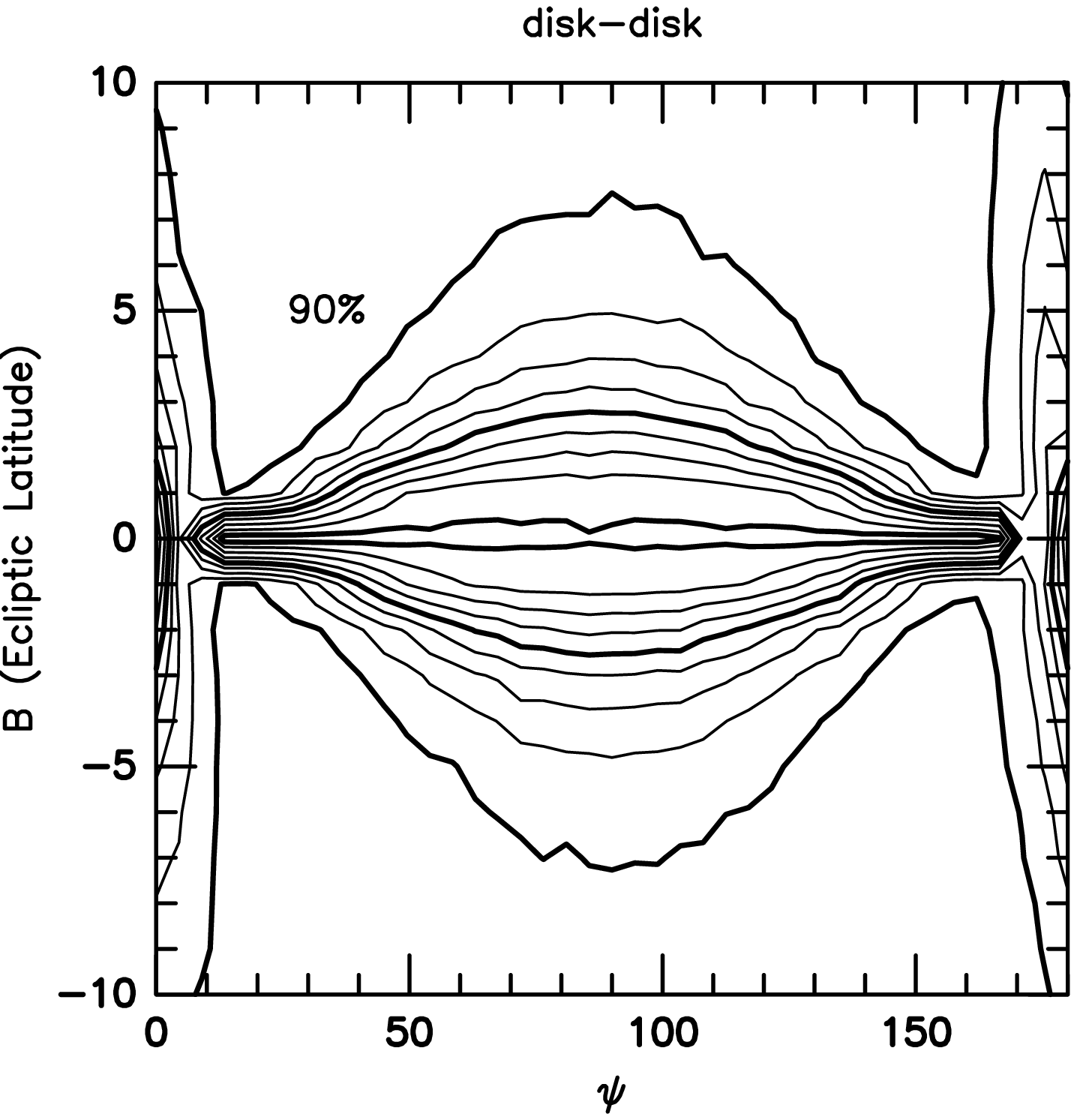}{6.4}{6.0}{6.4}
{
DBF as a function of $\psi$ and
${\rm{B}}$ for disk-bulge events. 
Contours are the same as those in Fig.\ {\three}.
}
\endinsert

Figure {\five} represents the DBF as a function of $\psi$ and $\rm B$, the
ecliptic latitude (${\rm B} = \alpha - 90^{\circ}$), for disk-bulge events.  
For this figure, we hold the ecliptic longitude fixed at 
 $\Lambda = 270^{\circ}$, and
vary ${\rm B}$ in the range 
$-10^{\circ} \leq {\rm B} \leq 10^{\circ}$.  In galactic coordinates, this
corresponds to the range 
$(l,b)=(-2^{\circ},-5^{\circ}) - (16^{\circ},4^{\circ})$.
For Baade's window, ${\rm B} = -6^{\circ}$, the DBF is greater than
90\% for the entire orbit.  Note that for ${\rm B} = 0$, it is impossible to 
break the degeneracy.  This is due to the fact that, at ${\rm B} = 0$, the
constraint given in equation \consta\ becomes
$\sum_{i=i}^{3}\alpha_i a_i = \Delta \omega + \Omega_{\oplus} 
\omega\Delta t \cot{\psi}  = 0$, 
and thus does not constrain the value of $\Delta \beta$
at all.  Similarly, for small values of $|{\rm B}|$, the value of 
$\Delta \beta$ is poorly constrained.  Thus, there is range of ecliptic
latitude,
 $|{\rm B}| < 1^{\circ}$, where the DBF is less than $40 \%$ 
for a majority of the orbit; this range should be avoided when
making observations.  Lines of sight where $|{\rm B}| > 2^{\circ}$ are 
optimal.

\topinsert
\mongofigure{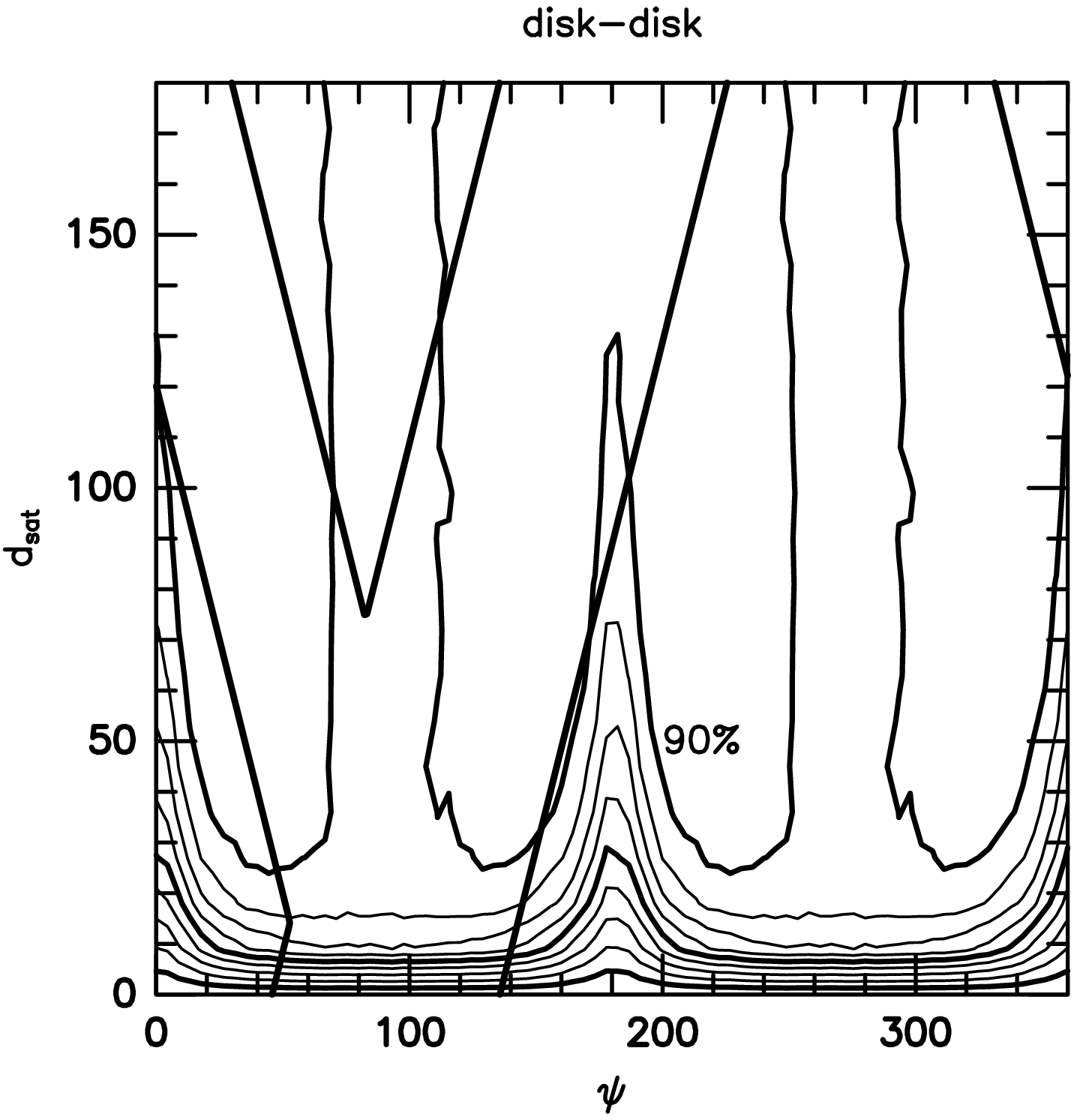}{6.4}{6.0}{6.4}
{
DBF as a function of $\psi$ and
$\dsat$ for disk-bulge events. 
 The y-shaped area of the graph between the heavy lines 
represents those range of $\psi$ for each value of $\dsat$ where 
observations are impossible due to the position of the Sun.  The other
contours are the same as those in Fig.\ {\three}.
}
\endinsert

\FIG\six{
DBF as a function of $\psi$ and
$\dsat$ for disk-bulge events. 
The y-shaped area of the graph between the heavy lines 
represents the range of $\psi$ for each value of $\dsat$ where 
observations are impossible due to the position of the Sun.  The other
contours are the same as those in Fig.\ {\three}.
}

Figure {\six} represents the DBF as a function of $\psi$ and $\dsat$, the 
Earth-satellite separation, for disk-bulge events.  Also shown is the range of $\psi$ for
which observations are impossible due to the position of the Sun.  This
range is a function of $\dsat$; thus by considering simultaneously
how this range
and the DBF vary as function of $\dsat$, 
we can find an optimal Earth-satellite separation.  For 
$\dsat \leq 30^{\circ}$, the range where observations are impossible 
is smallest ($\sim 90^{\circ}$), but the DBF is never greater than 80\%.  
For $\dsat = 60^{\circ}$, the observationally forbidden range is only 
slightly larger ($\sim 130^{\circ}$), and the DBF is greater than 90\% for
the majority of the orbit.  For $\dsat > 75^{\circ}$,
 the forbidden
ranges for the Earth and the satellite begin to diverge, and the times of
the orbit where observations are possible become separated.  Thus, although
the DBF continues to improve as $\dsat$ increases, there is an increased 
likelihood for these ranges that an observation will be foreshortened
due to the Earth or the satellite moving behind the Sun.  We consider 
$20^{\circ} - 75^{\circ}$ to be the optimal
range for the Earth-satellite separation. For $\dsat < 20^{\circ}$, the
DBF is unacceptably low.

We also considered the dependence of the DBF on the
parameters $\beta$, $M$, $\rm B$, and $\dsat$ for bulge-bulge events.  
We find the results to be almost identical to those for disk-bulge events, 
with variations of at most $\sim 10\%$, when the phase of the orbit 
is $\psi \sim 0$ or $180^{\circ}$, but typically the variations are much smaller.  
For reasons discussed above, the 
DBF is, in general, somewhat higher when $\psi \sim 90^{\circ}$, and somewhat
lower when $\psi \sim 0$ for bulge-bulge events.  As with disk-bulge events,
we find that there is no natural cutoff value for
$\beta$; for our fiducial value of $M=0.3 M_{\odot}$, 
the DBF is $\sim 90\%$ for the majority of the orbit; 
for Baade's window, ${\rm B} = -6^{\circ}$, the DBF is greater than
90\% for the majority of the orbit; and the optimal range of $\dsat$ is  
$20^{\circ} - 75^{\circ}$. 
Overall, the differences between bulge-bulge and disk-bulge events 
are minor, and will not significantly affect the
observations.  

\chapter{The Velocity Degeneracy}

In the previous section, we discussed the observational requirements of
a satellite to break the two-fold degeneracy in the projected
speed, $\tilde v$.  While measuring the projected speed is helpful in 
distinguishing between component populations, a measurement of the 
projected velocity, ${\bf{\tilde v}}$,
would provide a more conclusive distinction.

In this section we present the results for breaking the four-fold velocity
degeneracy, adopting the fiducial parameters used in the previous section:
$N_{meas} = 20$, $(\sigma_E,\sigma_S)=(1\%,2\%)$, $M=0.3M_{\odot}$,
 $d_{sat}= 60^{\circ}$ 
(2 months), $\beta_{max}=0.7$, and $\alpha=84^{\circ}$.  We calculate the
difference between the allowed solutions using equation \vecdif, and
the fractional error in each allowed solution using equation \veerr.  
As in the previous section, if the intrinsic errors in all allowed solutions
is $< 10\%$, and if the difference between the allowed solutions is 
$< 20 \%$ of the true solution, we consider
the degeneracy to be broken.   
Here we present the results for both disk-bulge and
bulge-bulge type events.

\topinsert
\mongofigure{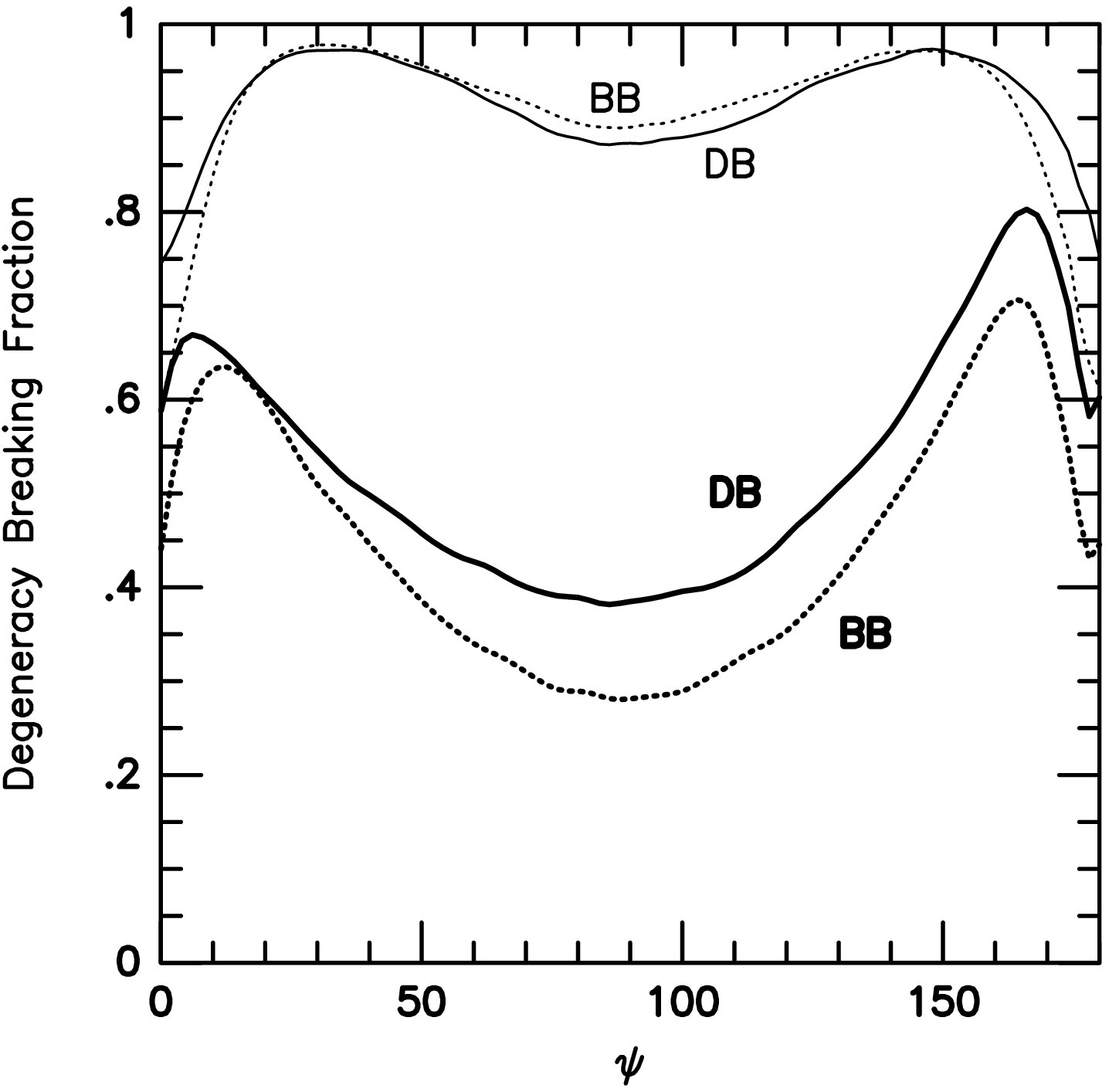}{6.4}{6.0}{6.4}
{
Vector DBF as a function of $\psi$ for
disk-bulge events (bold solid curve) and for bulge-bulge events (bold dashed curve).  
Also shown is the corresponding scalar DBF.  All other
parameters are fixed at $n_{obs} = 20$, $(\sigma_E,\sigma_S)=(1\%,2\%)$, 
$M=0.3M_{\odot}$, $d_{sat}=60^{\circ}$, $\beta_{max}=0.7$, and
$\alpha=84^{\circ}$.
}
\endinsert

\FIG\seven{
Vector DBF as a function of $\psi$ for
disk-bulge events (bold solid curve) and for bulge-bulge events (bold dashed curve).  
Also shown is the corresponding scalar DBF.  All other
parameters are fixed at $n_{obs} = 20$, $(\sigma_E,\sigma_S)=(1\%,2\%)$, 
$M=0.3M_{\odot}$, $d_{sat}=60^{\circ}$, $\beta_{max}=0.7$, and
$\alpha=84^{\circ}$.
}

Figure {\seven} represents the vector
DBF as a function of $\psi$ for disk-bulge events and bulge-bulge events.  
The scalar DBF is also 
shown for comparison.  We find that for disk-bulge events the DBF is greater than $\sim 40\%$ for
the entire orbit, and greater than $\sim 55\%$ for about half the orbit. 
 We find that for bulge-bulge events
the vector DBF is $\gsim 30\%$ over the entire orbit, and
$\gsim 40\%$ over half the orbit.  
In general, the analysis of the vector DBF curve 
is the same as that for scalar DBF curve:  there are minima at $\psi = 0$
and $180^{\circ}$ where the intrinsic errors dominate the DBF, and a minimum at
$\psi = 90^{\circ}$ where the ability of the constraint to distinguish between
the degenerate solutions dominates the DBF.  There are two unique features
of the vector DBF, however.  The first is that the minimum value of the DBF
occurs at $\psi \sim 90^{\circ}$, where the limiting factor is the ability 
to distinguish between the solutions.  In the scalar case, distinguishing
between the $\pm \Delta \beta_{-}$ solution
is unimportant, due to the fact that these
solutions have essentially the same projected speed.  To break the vector degeneracy, however,
one must distinguish between the $\pm \Delta \beta_{-}$ solutions, which is
often difficult when $\psi \sim 90^{\circ}$, where the constraint is the least
effective.  We also find a significant asymmetry in the vector DBF curve around
$\psi = 90^{\circ}$, the largest asymmetry being a difference of
$\sim 15\%$ between the DBF at $\psi = 15^{\circ}$ and $165^{\circ}$.  This
asymmetry is caused by fact that lenses in the disk have a preferred direction
of motion, and, to a lesser extent,
from the motion of the Sun around the galactic center.
For a given event, $\Delta \beta \propto \tilde v_\perp$, where
$\tilde v_{\perp}$ is the component of the transverse velocity perpendicular to the projected
Earth-satellite separation, ${\bf{r}}$.
As $\psi$ changes, the direction perpendicular to ${\bf{r}}$ changes with 
respect to the preferred direction of disk.  
Thus a preferred direction for ${\bf{\tilde v}}$ will
translate into an asymmetry in
the distribution of values of $\Delta \beta$ around $\psi = 90^{\circ}$.
Specifically, the values of $\Delta \beta$ are typically smaller when 
$\psi < 90^{\circ}$ than when $\psi > 90^{\circ}$.  Therefore, because the
$\pm \Delta \beta_{-}$ solutions are, on average, more closely spaced when
$\psi < 90^{\circ}$, the constraint is less likely to distinguish between them,
and the DBF is smaller.  This asymmetry does not arise in the scalar DBF curve
because it is not necessary to distinguish between the
$\pm \Delta \beta_{-}$ solutions. 
Note the smaller asymmetry around $\psi = 90^{\circ}$ for bugle-bulge events.
This is due to the fact that, in bulge-bulge events,
there is no preferred direction of motion of
the lens as there is in disk-bulge events.

{\bf Acknowledgement}: We would like to thank C. Han for making
helpful suggestions.
This work was supported in part by grant AST 94-20746 from the NSF and by
grant NAG5-3111 from NASA.

\endpage

\ref{Alcock, C. et al.\ 1995a, ApJ, 445, 133}
\ref{Alcock, C. et al.\ 1995b, ApJ, 449, 28}
\ref{Alcock, C. et al.\ 1995c, ApJ, 454, L125}
\ref{Alcock, C. et al.\ 1996a, ApJ, submitted (astro-ph=9512145) }
\ref{Alcock, C. et al.\ 1996b, ApJ, submitted (astro-ph=9606012) }
\ref{Alard, C.\ 1996, in Proc. IAU Symp. 173, Astrophysical Applications of Gravitational Lensing, 
ed. C. S. Kochanek \& J.N.Hewitt,
(Dordrecht: Kluwer), 215}
\ref{Ansari, R., et al.\ 1996, A\&A, in press}
\ref{Aubourg, E., et al. 1995, A\&A, 301, 1}
\ref{Bahcall, J. N., Flynn, C., Gould, A., \& Kirhakos, S.\ 
1994, ApJ, 435, L51}
\ref{Boutreux, T., \& Gould, A.\ 1996, ApJ, 462, 705}
\ref{Gould, A.\ 1994, ApJ, 421, L71}
\ref{Gould, A.\ 1995a, ApJ, 441, L21}
\ref{Gould, A.\ 1995b, ApJ, 441, 77}
\ref{Gould, A.\ 1996, PASP, 108, 465}
\ref{Gould, A., Bahcall, J. N., \& Flynn C.\ 1996, ApJ, 465, 759}
\ref{Gould, A. \& Welch, D. 1996, ApJ, 464, 212}
\ref{Griest, K.\ 1991, ApJ, 366, 412}
\ref{Han, C. \& Gould, A.\ 1995, ApJ, 447, 53}
\ref{Loeb A. \& Sasselov, D.\ 1995, ApJ, 449, L33}
\ref{Maoz, D., \& Gould, A.\ 1994, ApJ, 425, L67}
\ref{Mihalas, D., \& Binney, J., 1981, Galactic Astronomy,
(New York: W. H. Freeman)}
\ref{Nemiroff, R. J., \& Wickramasinghe, W. A. D. T.\ 1994, ApJ, 424, L21}
\ref{Paczy{\'n}ski, B.\ 1986, ApJ, 304, 1}
\ref{Udalski, A. et al.\ 1994, Acta Astron, 44, 165} 
\ref{Witt, H.\ 1995, ApJ, 449, 42}
\ref{Witt, H. \& Mao, S.\ ApJ, 430, 505}

\endpage

\APPENDIX{A}{Generalized Satellite Orbits}

For our Monte Carlo simulations, we assumed that the satellite is in an
Earth-like orbit, and that both the orbit of the Earth and the orbit of the
satellite are circular.  In reality, a satellite launched from the Earth will
not be in an Earth-like orbit, but will actually be in an elliptical orbit.
This implies that the true velocity of the satellite will differ from the
assumed velocity for each value of the Earth-satellite separation.  In this
appendix, we show that, for a reasonably small deviation from a circular orbit,
the difference between the true and assumed Earth-satellite velocity will
be small, and that therefore our results
are sufficiently generic to be applicable to a large class of orbits.

When launched from the Earth, the satellite will have some velocity with 
respect to the Earth.  We assume the direction of the satellite's 
velocity is parallel to that of the Earth (i.e. the orbit of the satellite
is in the plane of the Earth's orbit, and the satellite's perihelion is
at the distance of the Earth).  That is,
$\vsat = \ve + \delta v$.
We assume that the orbit of the Earth is 
circular, therefore 
a non-zero value of $\delta v$ implies that the satellite will
have an elliptical orbit with a period different from that of the
Earth's, and it will drift away from the Earth over time.
Using Kepler's laws, one finds that the
difference between the period of the Earth and the satellite is given by,
$$
{{\delta P}\over{P_{\oplus}}} = {{3\delta v}\over{\ve}}. \eqn\perio
$$ 
In \S\ 8.2, we determined
that the optimal Earth-satellite separation was in the range 
$20^{\circ} - 75^{\circ}$, or ${{2}\over{3}}
 {\rm{month}} - 2{{1}\over{2}}{\rm{months}}$.  We would like the 
Earth-satellite separation to vary in this range over the course of $\sim 3\,
{\rm{years}}$.  For definiteness, we choose the Earth-satellite separation
to be 3 months after 3 years, i.e. $\delta P/P_{\oplus} \sim 1/12$.
Thus, from equation \perio, 
and using $\ve = 30 \kms$ we find $\delta v \sim 0.8 \kms$. 
The orbit is nearly circular, with an eccentricity 
of $e = 0.06$.

\topinsert
\mongofigure{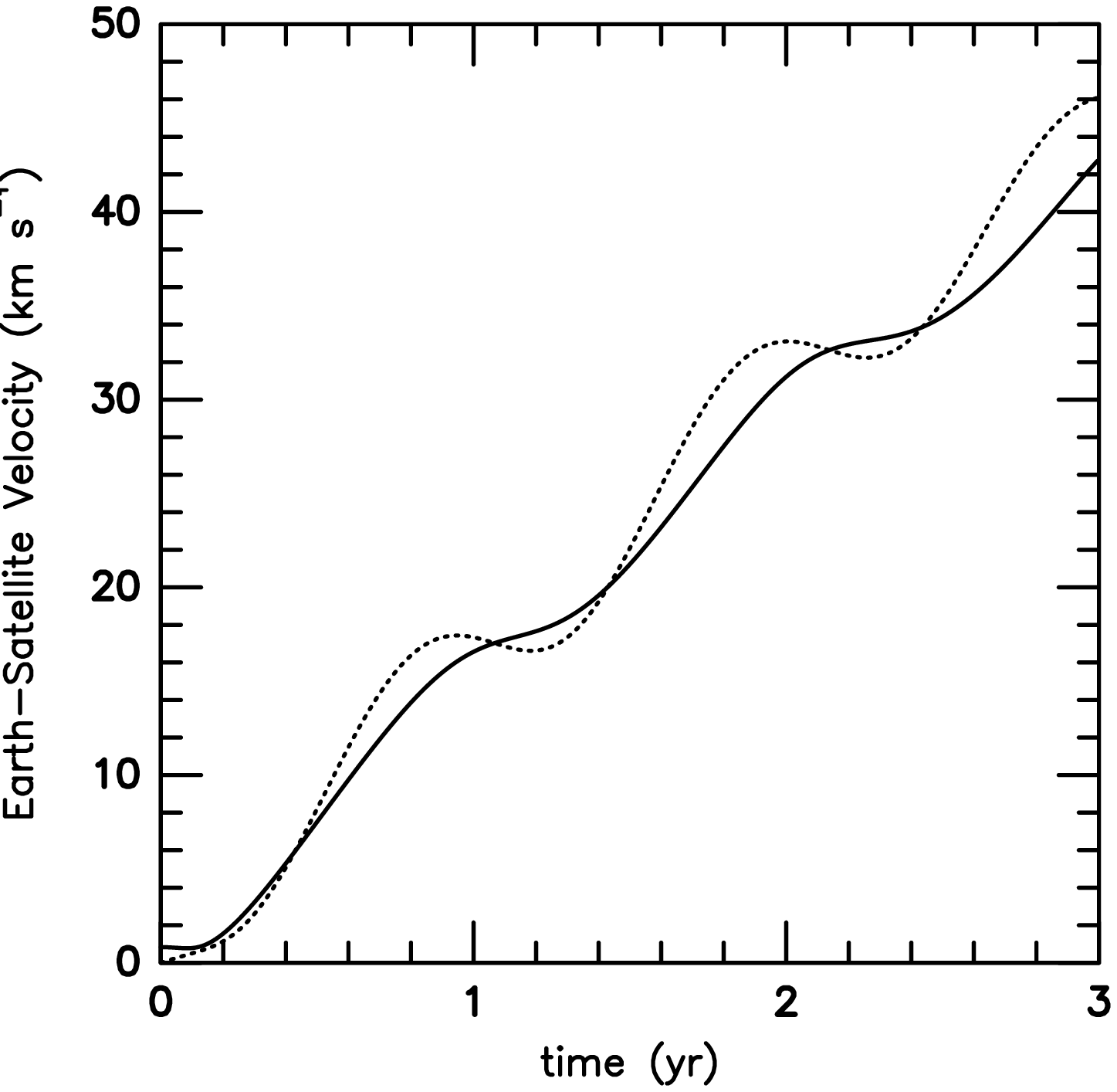}{6.4}{6.0}{6.4}
{
 Earth-satellite velocity as a function of time from launch.  
Shown are the Earth-satellite velocity assuming a circular orbit (dotted
line), and the real Earth-satellite velocity (solid line).  
}
\endinsert

\FIG\eight{
Earth-satellite velocity as a function of time from launch.  
Shown are the Earth-satellite velocity assuming a circular orbit (dotted
line), and the real Earth-satellite velocity (solid line).  
}

Using the parameters of the orbit, we can now calculate the difference between
the true value of the Earth-satellite velocity and the value used in our
simulation (which assumed a circular, Earth-like orbit for the satellite).  We
set $t=0$ as the time of launch, and at
each point in time, we calculate the true Earth-satellite separation and
velocity.  
We then calculate what the value of the Earth-satellite velocity would be
if the satellite were in an Earth-like orbit, separated from the Earth by
an amount equal to the true Earth-satellite separation.  This corresponds
to the velocity used in our Monte Carlo code.  Figure
{\eight} shows the Earth-satellite velocity for both cases in the range
$0 \leq t \leq 3\, {\rm{years}}$.  
Focusing on $t > 0.5\, {\rm{year}}$, we find 
that the true velocity is systematically smaller than the assumed velocity
by $\sim 6\%$, but varies from being $\sim 18 \%$ smaller to $\sim 6\%$ 
larger.

Here we have calculated the difference between the true Earth-satellite 
velocity and that assuming an Earth-like orbit for the satellite.  For the
purpose of degeneracy breaking, however, it is the projected Earth-satellite
velocity that is important.  In the case of an elliptical orbit, 
the magnitude of the projected Earth-satellite
velocity is dependent on the time of year in which the satellite is launched.
Since we do not know {\it{a priori}} when this will be, we cannot find 
precisely the difference between the true and assumed 
projected Earth-satellite velocity.  This is not strictly necessary, however, 
since the 
difference between the projected velocities will be of the same order as the
difference between the space velocities.  Since the ability to break
the degeneracy is roughly proportional to the projected Earth-satellite 
velocity, we conclude that the error introduced in our simulation 
by assuming an Earth-like 
orbit for the satellite is of order the error in the 
Earth-satellite velocity, $\sim 6\%$.

We conclude that the results obtained in our simulations
are applicable to likely satellite orbits.
If, however, one would like to consider satellite orbits
that are either highly inclined to the Earth's orbit, or highly eccentric, 
then our results are not applicable.  These situations will need to be 
analyzed individually.

\endpage
%\figout

\refout	
\endpage
\endpage
\endpage
\bye